\documentclass[10pt]{article}
\usepackage{amsmath,amssymb,amsthm}
\usepackage{graphicx}                        
\usepackage{caption}                        
\usepackage{float}                        
\usepackage{ucs}                        
\usepackage[utf8x]{inputenc}            
\title{\vskip -1.0 cm \bf Scintillation light simulation in big-sized BaF$_2$ and pure CsI crystals.\vskip 1.0cm }
\author{ \bf  \Large Z. Usubov   \\
\\
Joint Institute for Nuclear Research, Dubna, Russia    }

\theoremstyle{plain}

\begin{document}

\maketitle
\setlength{\parindent}{0pt}
\begin{abstract}
\noindent

We have investigated scintillation light  distribution in BaF$_2$ and pure CsI crystals   
with dimensions 3x3x20\,cm${^3}$
using the Geant4 toolkit.
The diffuse wrapping  material is  selected as coating  for the crystals.
 The simulated cosmic muons and 105 MeV electrons are  used as beam particles.
The optical attenuation along the crystals   is   explored 
with the simulation data. 
We have demonstrated the impact of the crystal surface 
finish on the light  distribution at the  crystal end,
optical photon arrival time,  incidence angle distributions,  and 
optical attenuation for the studied crystals.
\end{abstract}
\large  {
\section{Introduction}
 
$\quad$Flavor changing by all neutral current interactions is strongly suppressed in the Standard Model.
The new physics scenarios - supersymmetry, extra dimensions, little Higgs, quark compositeness -
naturally allow and predict the charged lepton flavor violation at some level (see, e.g.,\cite{newph}).

$\quad$The aim of the $\mu\to e$ conversion experiments is to search for the coherent
conversion of the muons from muonic atoms 
to the electrons in the field of a nucleus through some new 
lepton flavor violation interactions.
The conceptual designs\cite{come,mu2e} of the $\mu\to e$ conversion 
experiments include  a   calorimeter  able
to measure the energy of the electrons with the resolution $<$5\% for 105\,MeV 
and the time resolution  $\sim$1\,ns to provide the
trigger signal and measure track positions in addition to the tracking chambers.
The calorimeter will consist of the $\sim$3x3\,cm$^2$   dense
crystals and are $>$10 radiation lengths long.
\begin{table}[]
\large {
\begin{center}
{ 
\begin{tabular}{||c||c||c||c||}     \hline \hline
  Crystal                      & NaI(Tl)   &CsI(pure)   & BaF$_{2}$\\ \hline \hline
  Density\,(g/cm$^3$)          & 3.67      & 4.51       & 4.89     \\ \hline 
 Melting  Point  (${^0}$C)     & 651       & 621        & 1280  \\ \hline        
 Radiation  Length\,(cm)       &2.59       &1.86        & 2.03   \\ \hline            
 Moli\`ere  Radius  (cm)       & 4.13      &3.57        & 3.10   \\ \hline           
 Interaction  Length\,(cm)     & 42.9      &39.3        & 30.7     \\ \hline            
 Refractive  Index             & 1.85      &1.95        &  1.50    \\ \hline           
 Hygroscopicity                & Yes       & Slight     & No          \\ \hline            
 Luminescence  (nm)(at  peak)  & 410       & 420(310)   &300(220) \\ \hline            
 Decay  Time  (ns)             & 245       & 30(6)      &650(0.6-0.8) \\ \hline            
 Light Yield(Brightness)$(\%)$ & 100       & 3.6(1.1)   & 36.0(4.10) \\ \hline            
 d(LY)/dT($\%/{^0}C$)          &-0.2       &-1.4        &-1.9(0.1)      \\ \hline \hline            
\end {tabular}
\caption{Useful characteristics\cite{maoo}  of dense crystals as a Mu2e calorimeter material.
The values correspond to  the slow or fast(in parentheses)  scintilation component.}
}
\end{center}
}
\end{table}

$\quad$ In this paper we present the results of the Geant4 Monte Carlo simulation of the optical
processes in square cross section BaF$_{2}$\cite{bafp} and CsI\cite{csip} crystals. 
Both of these crystals are considered as   candidates for the electromagnetic calorimeter of
the Mu2e experiment.
The next section  briefly  describes      the strategy of simulating optical
photons in a crystal using  the Geant4 toolkit.
The properties of the crystals, choice of the  optical model,  and 
crystal  surface finish  are described.
In Section 3 we give the optical photon   simulation results.
We have examined the photon distributions at the end of the BaF$_{2}$ and CsI
crystals.                                       
Optical attenuation along the crystals was studied.   
The photon  arrival time and  the incidence angle 
were  estimated.
We  compared the simulation data  with the polished and unpolished 
crystal surfaces.
We end with a  short conclusion in Section 4.
\begin{figure}[ht!]                             
{\hskip  1.0cm} \includegraphics[scale=0.65]{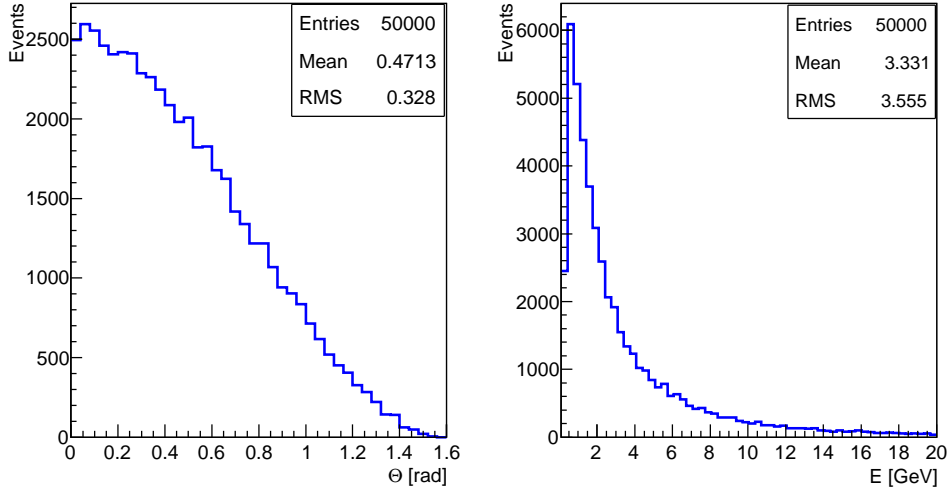}
\caption[]{The azimutal angular distribution (left panel) and energy spectrum (right panel) of
cosmic muons as simulated according to\cite{volk}.} 
\label{Norm}
\end{figure}

\section{ Some features of scintillating crystal modeling }

$\quad$Monte Carlo simulations play a crucial      role in 
determining the optimal crystal material and suitable calorimeter design. 
The Geant4 code  takes into account  optical properties of  materials  and
is a        reliable  tool for  studying a large class of scintillators.          

$\quad$ Two types of photons are involved in scintillation processes in  crystals:
high energy photons (e.g., 511 keV annihilation photons from  the $^{22}$Na or $^{68}$Ge/$^{68}$Ga, 662 keV
photons from  the $^{137}$Cs radioactive source, etc.) and low-energy optical photons (photons with a 
wavelength much greater than the typical atomic spacing).
The optical photons further undergo the following processes: bulk absorption, Rayleigh scattering,
reflection and refraction at  medium boundaries,  and  wavelength shifting.
The boundary processes on all crystal surfaces play  an important role in tracing  photons 
in crystals. Compared with them,  photon self-absorption  is less significant\cite{knoll}.

$\quad$ In Geant4\cite{gea4}  scintillator surfaces follow the
GLISUR (was realized earlier  for Geant3.21\cite{gea3}), 
LUT (look-up-tables)\cite{janec} 
 or UNIFIED\cite{unif} models.
The LUT model is based on measuring the angular reflectivity distribution inside  crystal 
and has been realized only for Bi$_{4}$Ge$_{3}$O$_{12}$(BGO) crystals as yet.  The
criterion  for BGO crystal size selection in these  measurements  is that   the 
largest surface length  must be 
smaller than 50 mm. This is an excellent way to model light transportation and boundary effects
in crystals, but it  
is not applicable to our  simulation.

$\quad$In the UNIFIED model the surfaces are made up of micro-facets with normal vectors that 
are oriented around the average surfaces according  to the Gaussian
distribution with mean 0 and
standard deviation given by a user-adjustable value and known as $\sigma_{\alpha}$.
The magnitude of this deviation determines whether  the
surface is polished, etched, or ground and is  set to 1.3$^{0}$, 3.8$^0$, and  12$^{0}$, respectively. 
These  options of crystal surface are combined with different
coating  conditions. In this simulation we use the UNIFIED model for the processes between two
dielectric materials (surface type was set to dielectric-dielectric).  
We combined the  polished and ground surface finishes with the 
frontpainted crystal wrapping option,  which implies the absence of an air gap between the crystal
and the wrapping and represents  diffuse (Lambertian) reflection.

$\quad$To simulate the whole process in the scintillators, the physics list included low-energy electromagnetic
physics    and     scintillation and transportation of  optical photons. The index of refraction and  the 
fast and slow components of the scintillation photon distributions 
for BaF$_{2}$ and CsI as a function of the  wavelength were used.
The scintillation photons are generated as  a pure Poisson process (RESOLUTIONSCALE parameter was set to 1).
The brightness of the
crystal is       important      for the energy and timing resolution of the calorimeter.
The corresponding number of  scintillation photons was  calculated in this analysis assuming that  brightness
of BaF$_{2}$ and CsI is 11.8 photons/keV and 2.8 photons/keV, respectively. 
The Birk's kB' constant was taken to be 0.00368\,mm/MeV and 0.00152\,mm/MeV for BaF$_{2}$
and CsI, respectively. The relative strength of the fast component as a fraction of total 
scintillation yield is given by the YIELDRATIO. This variable values 15$\%$ for BaF$_{2}$
 and 80$\%$ for CsI was used in simulation. The simulation
results for effective decay time was  determined to be
638.8\,ns and 12.5\,ns respectively  for BaF$_{2}$ and CsI crystals.       

\section{Simulation and results}

$\quad$  This simulation was  performed using    Geant4.10.0
for inorganic scintillators  BaF$_{2}$ and pure CsI 
with the dimensions
3x3x20$\,$cm$^{3}$. 
The properties of these crystals\cite{maoo}  are
compared with NaI(Tl) in Table$\,$1.
We also note that BaF$_{2}$ is the fastest scintillator now      and  pure CsI is 
quite soft and one of the 
cheapest crystals.  
For both crystals all surfaces are polished or ground. We collected photons  from one of the crystal ends
(3x3\,cm$^{2}$) (hereinafter  referred to as photodetector side). 
On the  photodetector side  of the crystal  photons are fully absorbed.
All other surfaces were wrapped in a highly reflective (R=98$\%$)  diffuse coating without 
an air gap.

$\quad$Cosmic muons were generated according to\cite{volk} in the range 0.3-5000\,GeV
and injected always perpendicular 
to the 3x20 cm$^{2}$ crystal surface. 
 The azimuthal angular 
distribution and energy spectrum of the simulated cosmic muons are shown in Figure\,1.

$\quad$Figures\,2 and 3 show the XY position-dependent number of optical photons 
as seen by the photodetector side   
of the polished and ground  BaF$_{2}$ and CsI crystals.
The  5000 cosmic muons   impinge   perpendicularly  on the crystal lateral surface
in -Y direction
at different distance l from  the photodetector side along  the crystal  Z-axis.
Scintillation  occurs    at various Z positions in a
crystal.         
It can be seen that if the muon impinging point is close to the crystal photodetector side,
the photon distribution  for  the  polished crystal  is 
not uniform and smoothed with increasing l.
These figures illustrate the difference of the light 
distribution for two crystals.                                           
The ground surface of the 
crystal leads to the focusing of the light on the center of the crystal photodetector side. 
On the other
hand,  if the surface  is  polished,  significantly more photons  reach the end of the 
crystals. 

$\quad$Figures\,4 and 5 plot the  mean  values and 
standard deviations  of the fit with 
the Gaussian   function of the
optical photon distributions on the  photodetector
side as a function of the beam impinging position, actually the position
of the deposited energy inside the crystal. The simulated data correspond to
the CsI (left panel) and BaF$_{2}$ (right panel) crystals.

$\quad$ As expected, the effective attenuation length $\lambda_{eff}$ decreases when 
the crystal faces are roughened. This is because some photons leave       the crystal
instead of being totally reflected. The path of the photons will be significantly changed. 
Note that the  total bulk attenuation length  $\lambda_{eff}$,
absorption attenuation length   $\lambda_{ab}$, and scattering attenuation length 
 $\lambda_{s}$  are related as

$$ {1\over {\lambda_{eff}}} = {1\over {\lambda_{ab}}} + {1\over {\lambda_{s}}}. $$ 

As shown in the figures
significant light loss is     caused by the unpolished crystal surface.    
With the                               
BaF$_{2}$ crystal,  more photons appears to be collected according to the higher 
brightness of crystal.

$\quad$ Figure\,6 show the XY position-dependent optical photons  at the photodetector side
of the polished and ground BaF$_{2}$ and CsI crystals when  2000 electrons with E=105 MeV
impinge   on the center of the opposite  side along  the Z-axis.
It can be clearly seen that the distributions are  uniform for  the    
polished crystal surfaces. The photons are more focused on the center of  the 
photodetector side if the crystal surface is     ground. These distributions have  
the same behavior as in the case of  cosmic muons impinging on the crystals.
The corresponding number of photon distributions is 
shown in Figure\,7. The results of the Gaussian fit are  also shown in figure.                                  
The decrease the number of  optical photons  due to the surface 
roughening is  $\sim$4.8 and $\sim$4.1
times for BaF$_{2}$ and CsI, respectively. 

$\quad$ We note that  the energy depositions in  the  3x3x20\,cm$^3$ 
crystals when 105 MeV electrons impinge perpendicularly on the 
center of the 3x3\,cm$^{2}$ side is  73.96 $\pm$ 0.06 MeV  and 70.33 $\pm$ 0.06 MeV 
for BaF$_{2}$ and CsI, respectively.

$\quad$ The photon arrival  time is important for the time resolution
of the detector\cite{span}.        
The dependence of the photon arrival  time on the distance of the traversing 
muon to   the photodetector side is shown in Figure\,8 for the BaF$_{2}$ crystal with
the ground and polished surfaces. 
It is seen  that for the polished  crystal (right column)
the distributions have  two peaks. The photons which
travel directly to the photodetector side without undergoing any optical interaction
 give the first peak. 
The photons that are reflected  from the opposite end of the crystal and arrive   
at the photodetector  side
without any other reflections 
give the second peak. The peak positions correspond to the        
distance from the beam impinging point to the photodetector side. The second peak
amplitude increases rapidly with decreasing distance from  the beam impinging point
to the opposite side.
In contrast, for the crystal with the 
ground surface only  the peak from  the photons impinging on  the photodetector side without any 
previous reflections  is clearly seen. Again,  the peak position corresponds  to the beam 
impinging point.                                                             
In this case the "indirect" photons (which undergo reflection from  the lateral sides)
lead to the  broadening of the distribution
with increasing distance to the photodetector side.
 
$\quad$ Figure\,9 demonstrates  the  arrival time and Figure\,10 the incidence
angle 
(the angle  with respect
to the crystal Z-axis)
distribution for optical photons at  the photodetector side             
for  the polished and ground BaF$_{2}$
and CsI crystals. The electrons with E=105 MeV 
impinge  on  the center of the  opposite side.
In Figure\,10,  -1 on X-axis corresponds to the normal incidence of photons.
The ratios of the mean photon arrival time 
in the CsI crystal to the mean photon arrival time in the  BaF$_{2}$  
crystal for the                              
polished and ground surfaces are 1.8  and 2.1, respectively. Note that
the ratio  between the refraction indices of CsI and BaF$_{2}$ is 1.3.
This simulation shows that the polished and ground crystal surfaces lead to  
different optical photon angular distributions.
The number of photons depends strongly on the incidence angle.                         
        
\begin{figure}[t]                             
\vskip -2.4cm \hskip 1.5cm \includegraphics[scale=0.9]{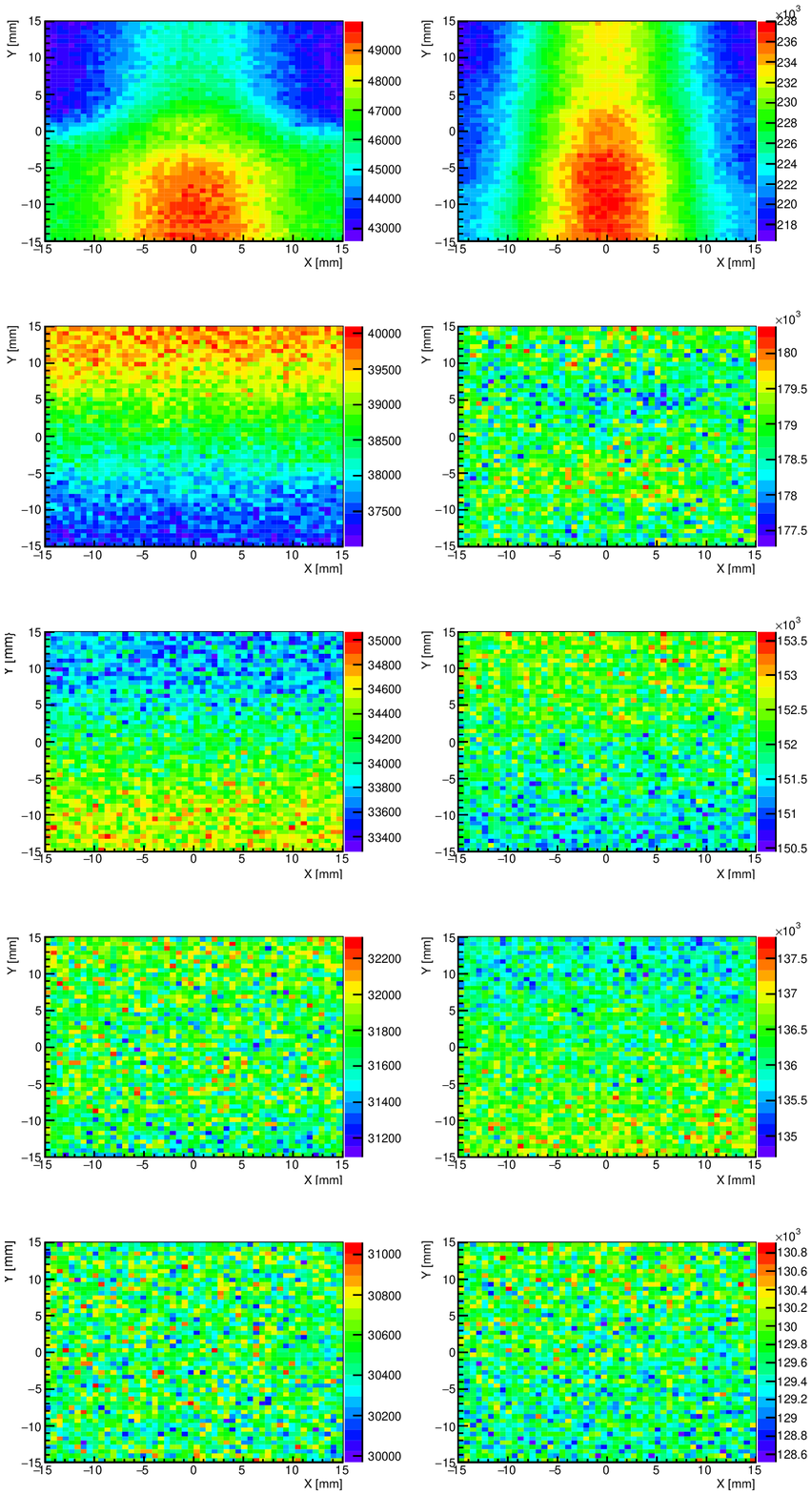}
\caption[]{ XY plot for optical  photons that reached the photodetector side  of the 
polished CsI (left column) and
BaF$_{2}$ (right column) crystals  with diffuse wrapping. Cosmic muons impinge   perpendicularly
to  lateral side at a  distance from the photodetector side 
of l = 2, 6, 10, 14,  and 18 cm (rows from top 
to bottom) in -Y  direction.  }  
\label{Norm}
\end{figure}

\begin{figure}[ht]                             
\vskip -2.5cm \hskip 2.0 cm \includegraphics[scale=0.90]{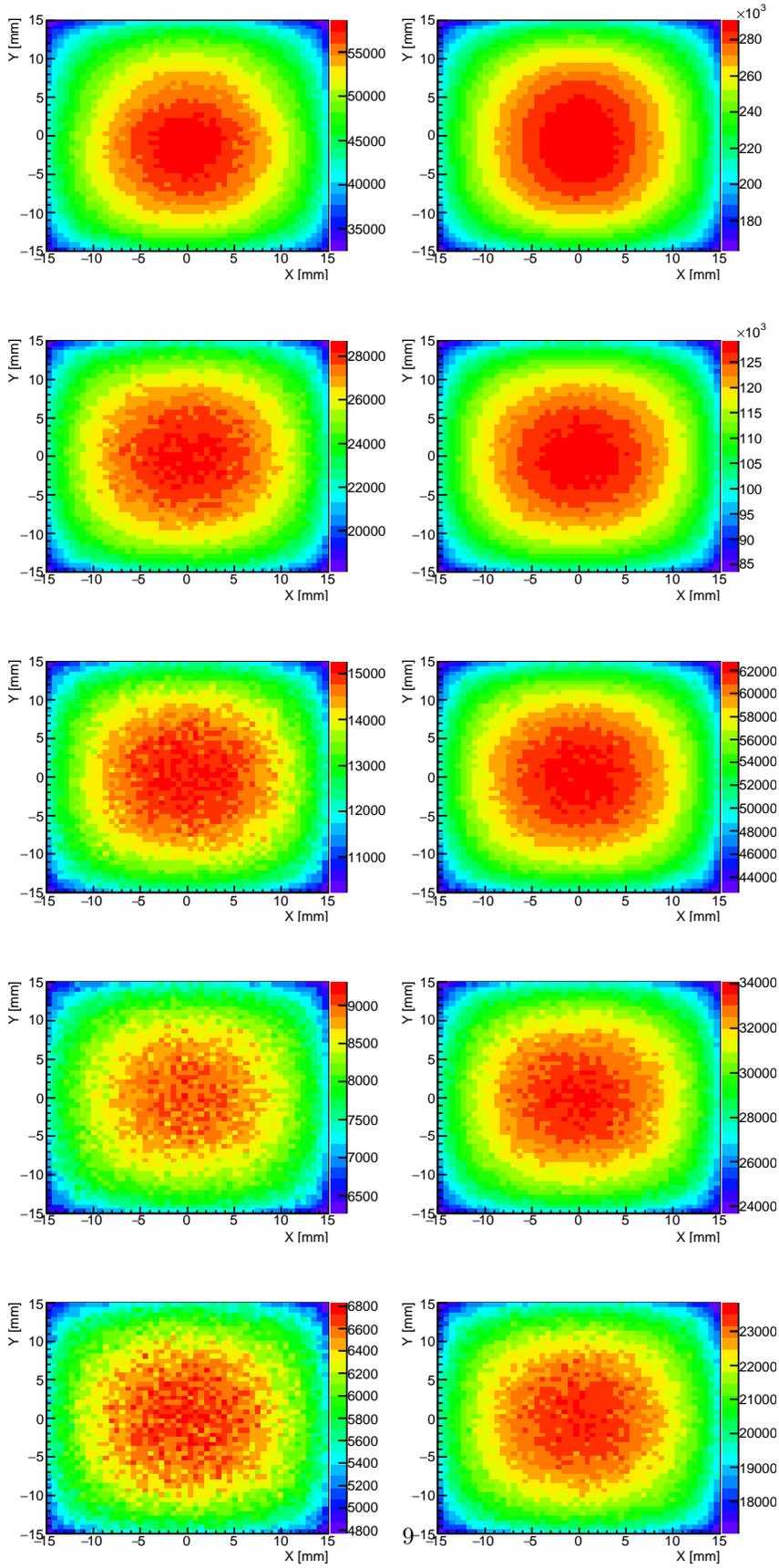}
\caption[]{ Same as in Figure\,1, but for the ground  crystal  surfaces.}         
\label{Norm}
\end{figure}

\begin{figure}[ht!]                             
\vskip  1.5cm \hskip  1.5cm \includegraphics[scale=0.65]{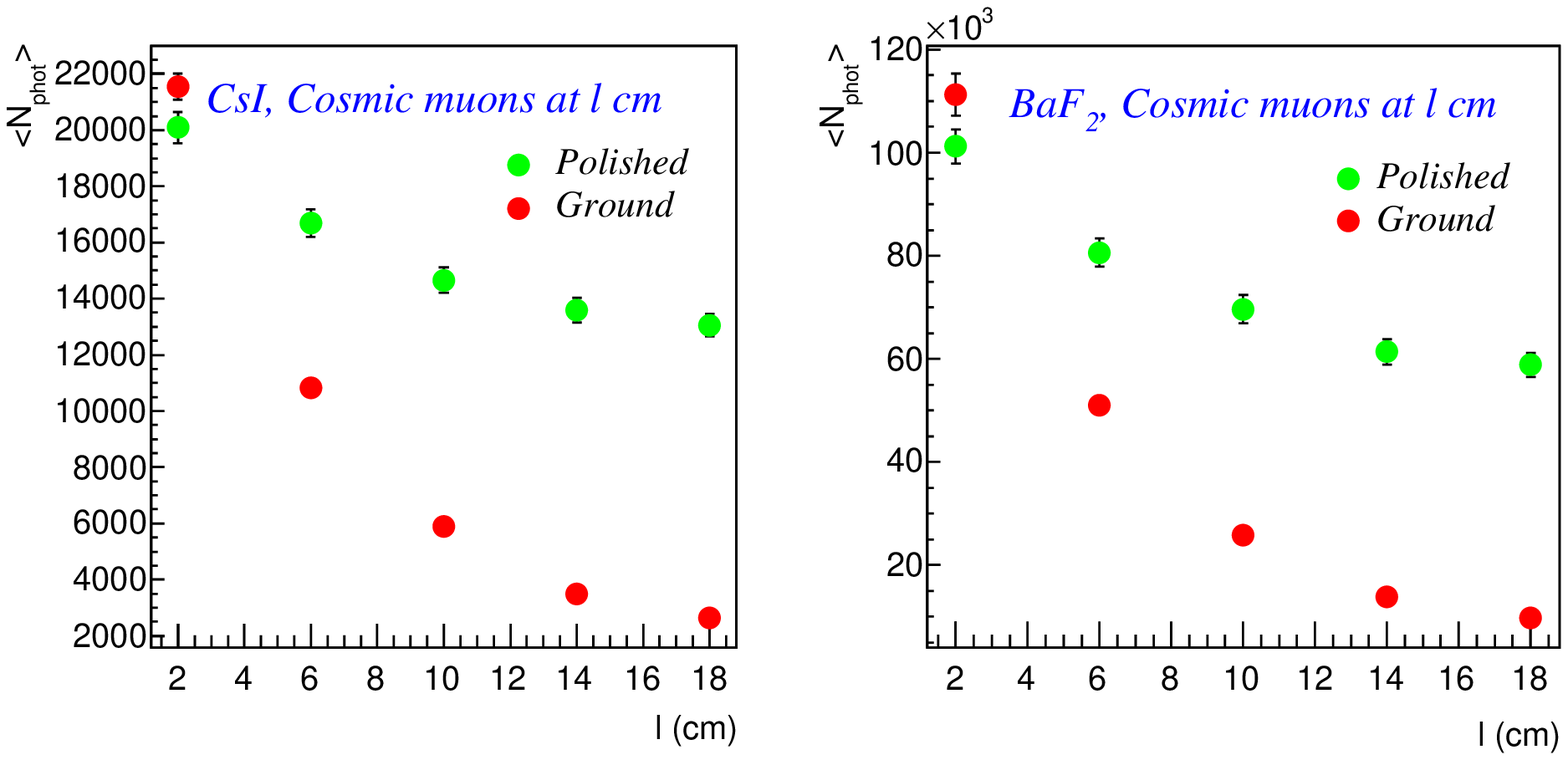}

\vskip -2.0cm \caption[]{ The mean   value 
of the Gaussian  fit of  the number of optical photons arriving  
at the photodetector side of the CsI (left panel) 
and  BaF$_{2}$ (right panel) crystals as a function of the distance from the 
cosmic muon impinging point.}
\label{Norm}
\end{figure}

\begin{figure}[ht!]                             
\hskip  2.0cm \includegraphics[scale=0.60]{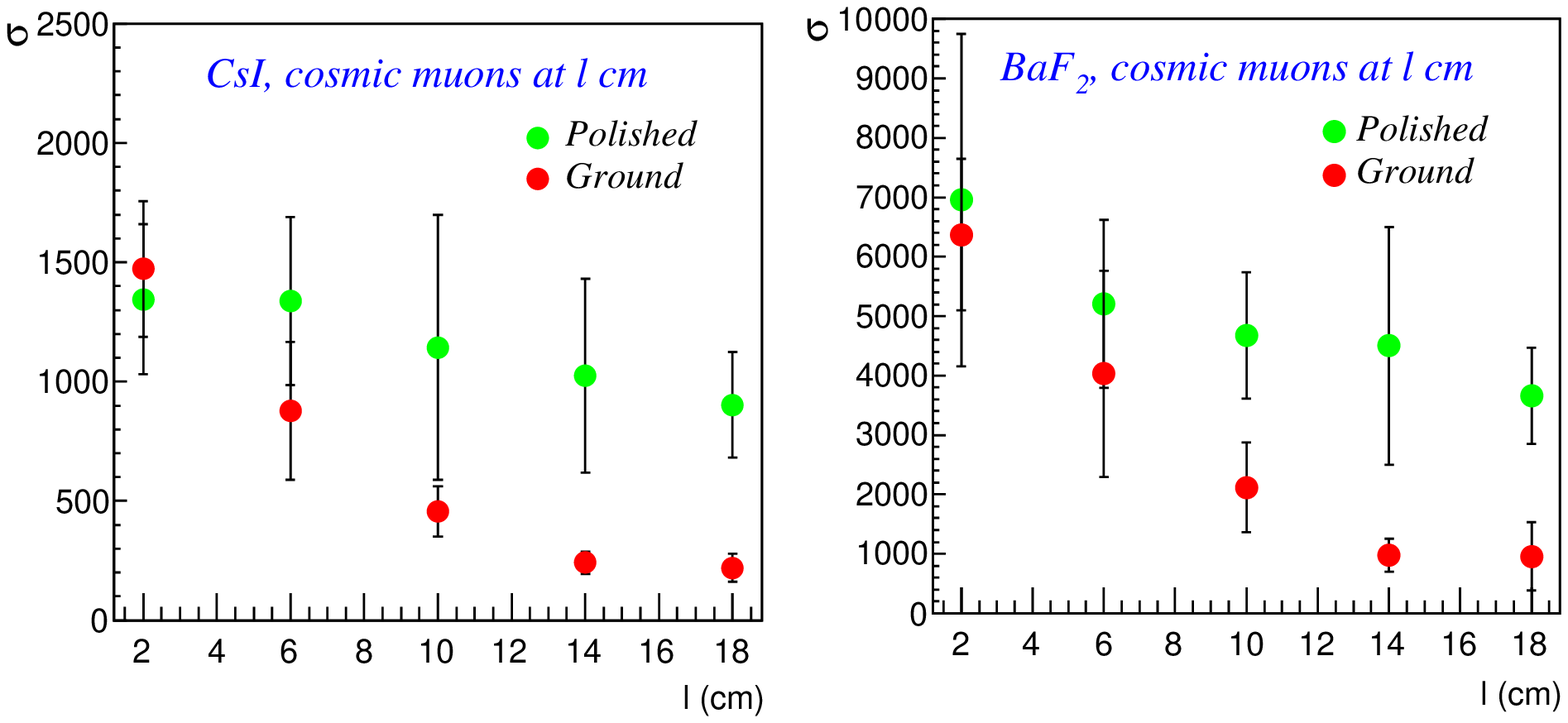}
\vskip -2.0cm \caption[]{The standard deviation of the 
 Gaussian  fit of the number of optical photons arriving at the photodetector side  
of CsI (left panel)
and BaF$_{2}$ (right panel) crystals as a function of the distance from the cosmic muon  impinging 
point.} 
\label{Norm}
\end{figure}

\begin{figure}[ht!]                             
{\hskip  1.5cm} \includegraphics[scale=0.65]{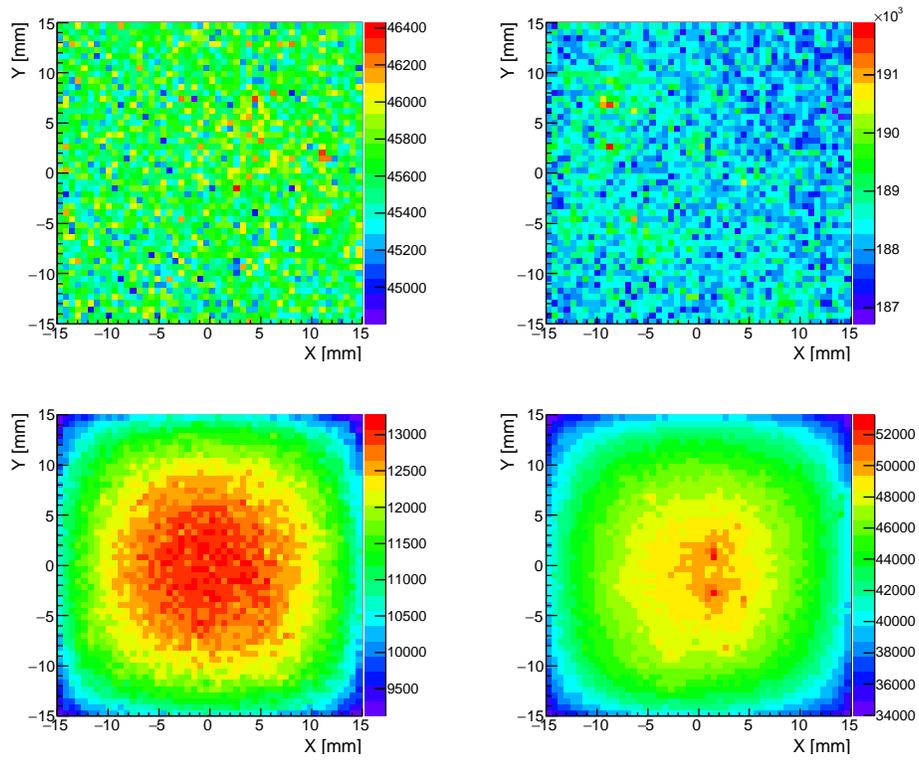}
\caption[]{XY plot for the optical photons that reached the photodetector side
of the CsI (left column) and BaF$_{2}$ (right column) crystals, 
polished (top row) and ground (bottom row). The E=105 MeV electrons  impinge  on  the 
center of the opposite
side of the crystal perpendicularly.}
\label{Norm}
\end{figure}

\begin{figure}[ht!]                             
{\hskip  1.5cm} \includegraphics[scale=0.60]{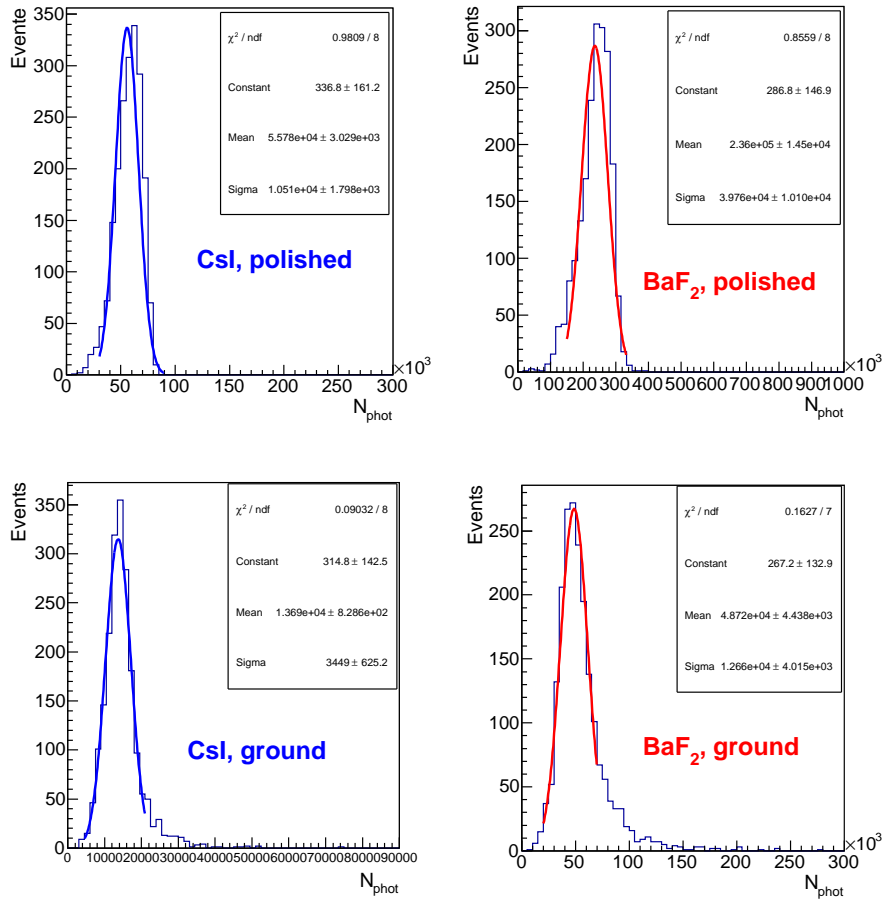}
\caption[]{The optical photon number distributions corresponding to the case shown 
in  Figure\,6.}

\label{Norm}
\end{figure}

\begin{figure}[ht!]                             
\hskip  1.0cm \includegraphics[scale=0.70]{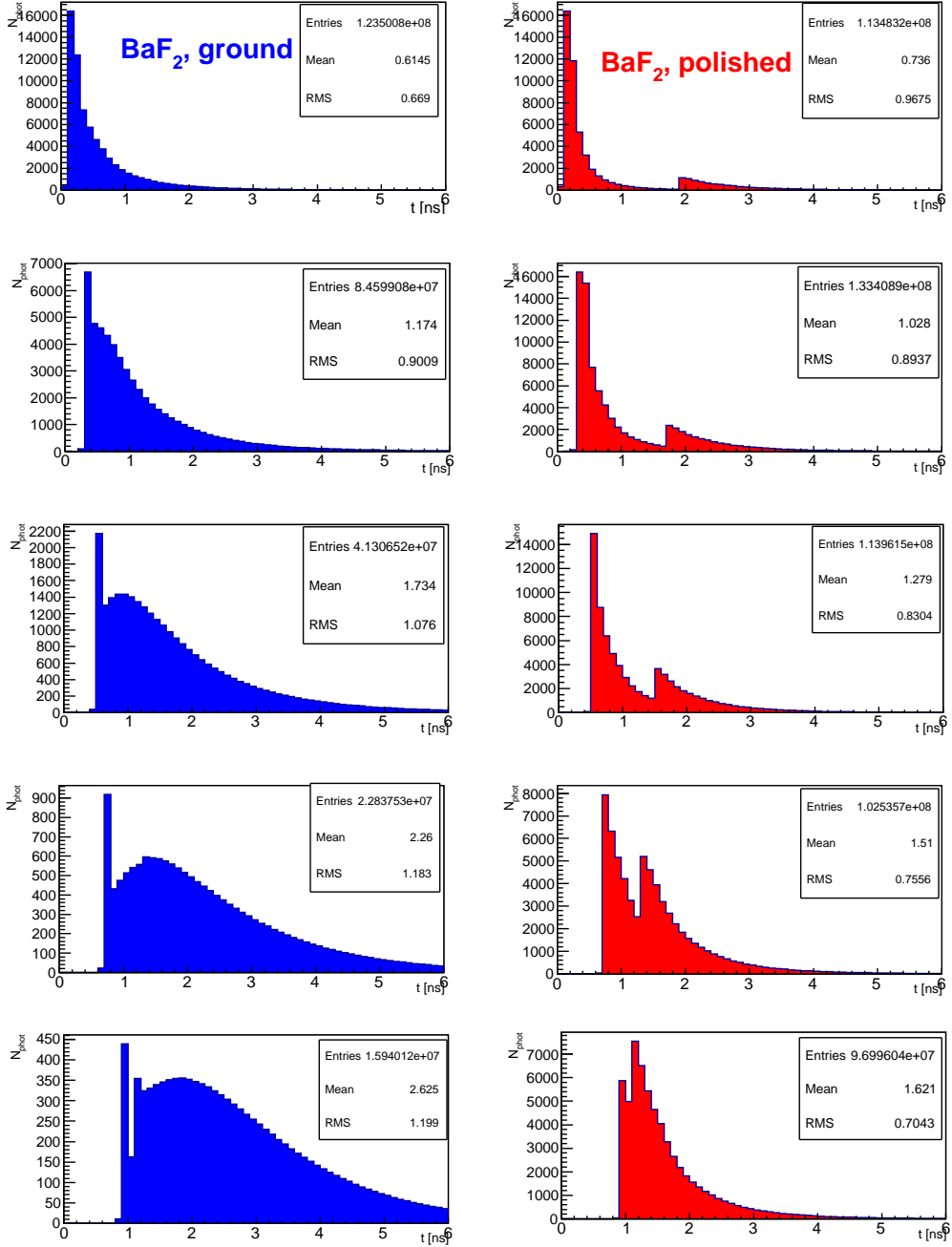}
\caption[]{Optical photon arrival  time at the photodetector side of the 
polished (right column) and ground (left column)   BaF$_{2}$ crystals. Cosmic muons
impinge  perpendicularly to the lateral 
side  at a distance of  l = 2, 6, 10, 14,  and 18 cm ( rows from top to bottom)
from the  photodetector side in -Y direction.}

\label{Norm}
\end{figure}

\begin{figure}[ht!]                             
\hskip  1.5cm \includegraphics[scale=0.60]{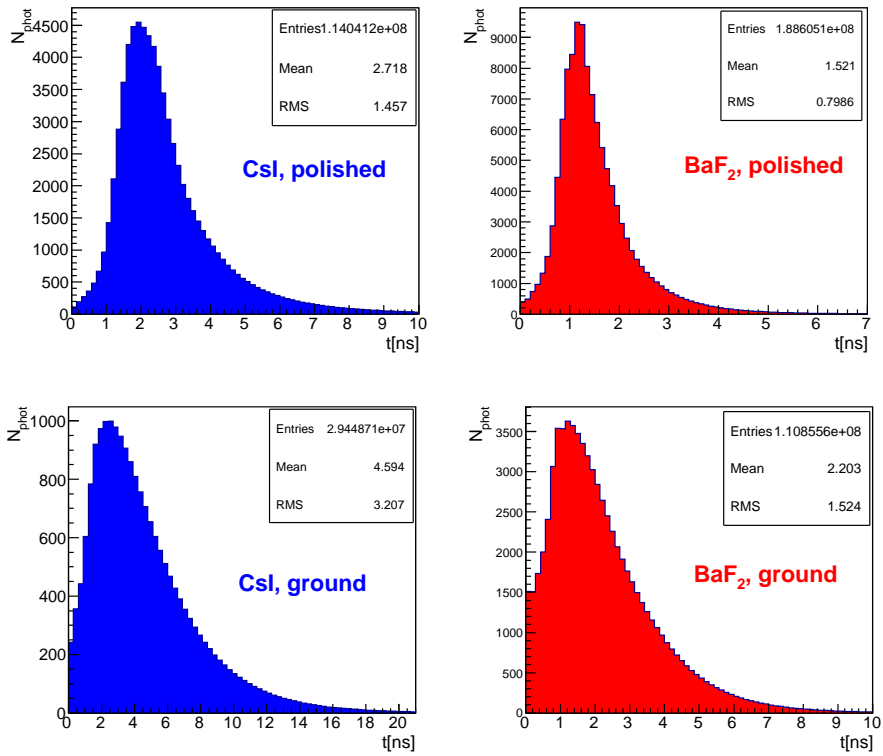}
\caption[]{Optical photon arrival  time at the photodetector side of the CsI (left column)
and BaF$_{2}$ (right column) crystals. The crystal surfaces are polished (top row) or ground (bottom row).
Electrons with E=105 MeV
impinge on the center of the opposite side perpendicularly.}               

\label{Norm}
\end{figure}

\begin{figure}[ht!]                             
\hskip  1.0cm \includegraphics[scale=0.66]{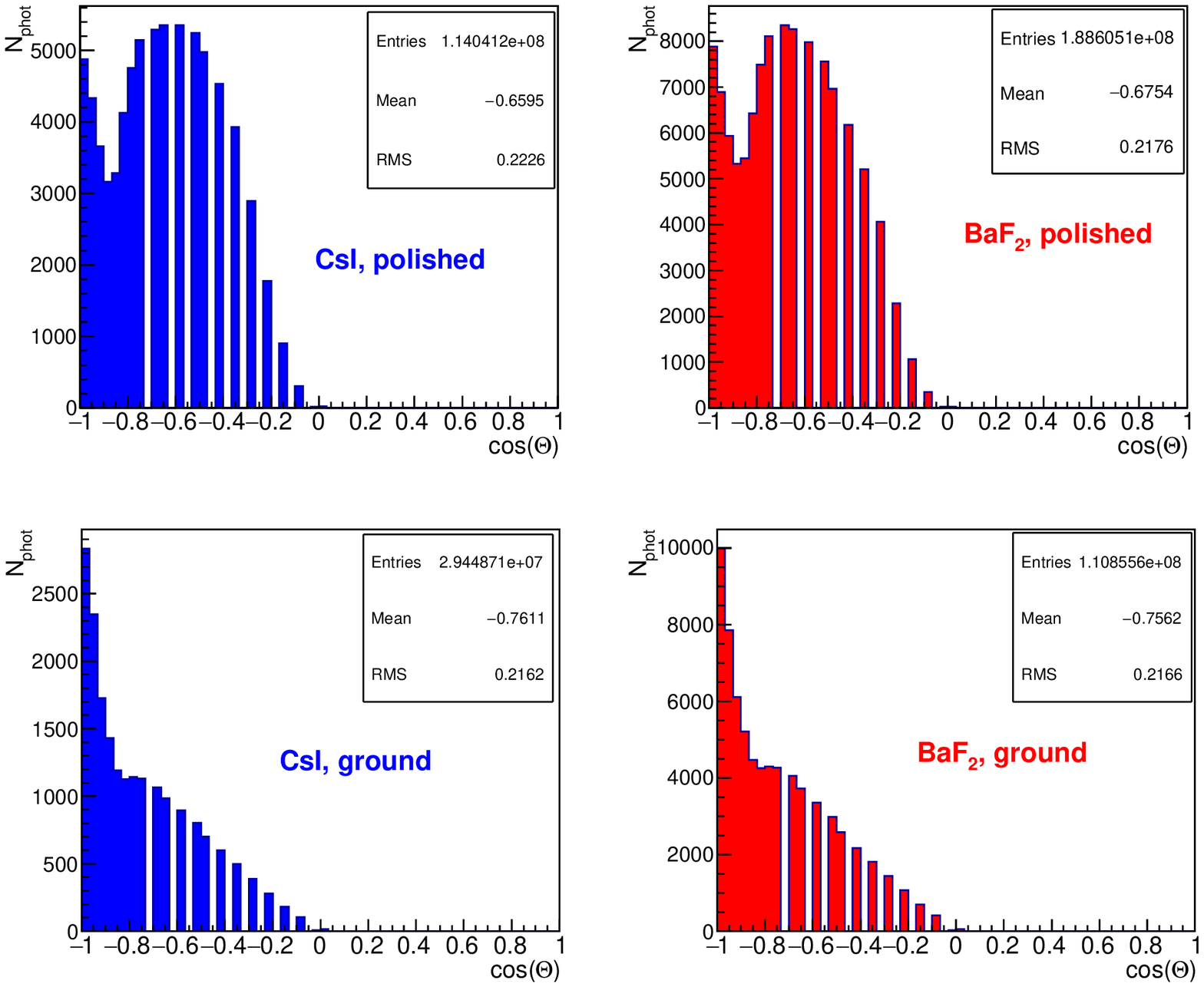}
\caption[]{Incidence angle distribution of the optical photons on the  photodetector side of
the CsI (left column) and BaF$_{2}$  (right column)  crystals. The crystal surfaces 
are polished (top row) or ground (bottom row).  Electrons with E=105 MeV
impinge on the center of the  opposite side perpendicularly.}               

\label{Norm}
\end{figure}

\section{Conclusion }

$\quad$ We presented the results of  the Geant4 simulation of 
the optical photon transport
and  surfaces border effects 
in the  3x3x20\,cm$^{3}$ BaF$_{2}$ and pure CsI crystals.
Cosmic muons and electrons with E=105 MeV
were  used as  beam particles. 
The effect of surface roughening on scintillation light
was studied in crystals with diffuse wrapping. 

$\quad$ We found that the crystal surface finish plays  a crucial role in the 
spatial distribution and absolute value of optical photons in scintillating crystals. 
The simulation studies  showed  that in  the crystals with the unpolished
surface the effective attenuation length decreases and hence
 polished  crystals make 
it possible to collect significantly more photons.         
However, the ground crystal surface gives more light focused on  the center
of the crystal end.

$\quad$ The photon arrival  time at the crystal photodetector side 
was explored as a function of  the 
cosmic muon traversing position and the  crystal surface treatment.
The impact of the crystal surface finish on the optical photon incidence angle 
was also demonstrated.

\vspace {0.8cm}


   $\quad$ We gratefully acknowledge the very helpful discussions and suggestions by
Y.\,Budagov, Y.\,Davydov, V.\,Glagolev and P.\,Murat.
}

\bibliographystyle{plain}
\bibliography{martingales}

\end{document}